\documentclass[a4paper,twocolumn,11pt,floatfix,accepted=2023-08-13]{quantumarticle}
\pdfoutput=1
\usepackage[utf8]{inputenc}
\usepackage[english]{babel}
\usepackage[T1]{fontenc}
\usepackage{amsmath}
\usepackage{hyperref}

\usepackage{tikz}
\usepackage{lipsum}

\usepackage{braket}
\usepackage{amsmath}
\usepackage{array}
\usepackage{tabularx}
\usepackage{mathtools}
\usepackage{commath}
\usepackage{bbold}
\usepackage{graphicx}
\usepackage{comment}
\usepackage{amssymb}
\usepackage{xcolor}
\usepackage[numbers]{natbib}

\begin{document}
\title{Modified dipole-dipole interactions in the presence of a nanophotonic waveguide}
\author{Mathias B. M. Svendsen}
\affiliation{Institut f\"ur Theoretische Physik, Universit\"at Tübingen, Auf der Morgenstelle 14, 72076 T\"ubingen, Germany}
\author{Beatriz Olmos}
\affiliation{Institut f\"ur Theoretische Physik, Universit\"at Tübingen, Auf der Morgenstelle 14, 72076 T\"ubingen, Germany}
\affiliation{School of Physics and Astronomy and Centre for the Mathematics and Theoretical Physics of Quantum Non-Equilibrium Systems, The University of Nottingham, Nottingham, NG7 2RD, United Kingdom}

\begin{abstract}
When an emitter ensemble interacts with the electromagnetic field, dipole-dipole interactions are induced between the emitters. The magnitude and shape of these interactions are fully determined by the specific form of the electromagnetic field modes. If the emitters are placed in the vicinity of a nanophotonic waveguide, such as a cylindrical nanofiber, the complex functional form of these modes makes the analytical evaluation of the dipole-dipole interaction cumbersome and numerically costly. In this work, we provide a full detailed description of how to successfully calculate these interactions, outlining a method that can be easily extended to other environments and boundary conditions. Such exact evaluation is of importance as, due to the collective character of the interactions and dissipation in this kind of systems, any small modification of the interactions may lead to dramatic changes in experimental observables, particularly as the number of emitters increases. We illustrate this by calculating the transmission signal of the light guided by a cylindrical nanofiber in the presence of a nearby chain of emitters.
\end{abstract}

\maketitle

\section{Introduction}
An ensemble of emitters coupled to a common environment displays collective behaviour. This includes the enhanced and inhibited emission of photons from the ensemble (so-called super and subradiance, respectively) \cite{PhysRev.93.99,Lehmberg-1970,James-1993}, and the emergence of induced dipole-dipole interactions among the emitters \cite{Sutherland-2016,Needham-2019,Damanet2016}. The rates associated with these processes can be modified via the specific choice of the environment and its boundary conditions, which may be tailored in order to obtain desirable properties. For example, introducing a dielectric or metallic structure in close proximity to an atomic gas, modifies the local electromagnetic spectrum and hence the collective behaviour of the ensemble \cite{Jones-2018,Sinha-2018,Fuchs-2018,PhysRevA.95.033818,PhysRevB.82.075427}. This effect was noticed by Purcell in 1950 \cite{PhysRev.69.37} and has later been verified in various experiments involving atoms or electrons in cavities, and molecules coupled to dielectric structures \cite{PhysRevLett.50.1903, PhysRevLett.55.67, PhysRevLett.55.2137, PhysRevLett.65.1877}.

Among these structures, so-called nanophotonic waveguides, such as single mode optical nanofibers \cite{PhysRevLett.104.203603,Ding:10} or integrated photonic nanostructures \cite{Lodahl2015,Ritter2018,Skljarow2020,Skljarow2022}, particularly stand out since they provide strong and homogenous coupling between the emitters and long coherence times \cite{Reitz2013}, and also due to the confined or guided field modes carried by these nanostructures, which can lead to propagation-direction-dependent (chiral) emission \cite{Lodahl2017}. The translationally invariant nature of these guided modes gives rise to infinitely ranged coherent dipole-dipole interactions and incoherent couplings between the emitters. These all-to-all interactions have recently facilitated, for example, the observation of super- and subradiance \cite{Pivovarov2021,Solano,Pennetta2022}, the study of non-trivial topology \cite{Vega2021,Bello2022,McDonnell2022}, the realization of long-lived photon storage and multiple photon bound states \cite{Asenjo2017,Albrecht2019,Buonaiuto2019,Mahmoodian2020,Wang2022} and the investigation of new dynamical phases and phase transitions \cite{Sedov2020,Buonaiuto2021}. The modes propagating in the space outside of the nanostructure are often referred to as radiation or unguided modes. Unfortunately, since these modes can propagate in any direction in space, their contribution to the coherent and incoherent interactions between the emitters is highly non-trivial, containing, for example, frequency integrals that either converge very slowly or not at all. Hence, they are often approximated by the free-space modes \cite{Solano,PhysRevA.95.023838}, an approximation that, however, inevitably breaks down when the emitters are close to the nanostructure. A number of works have nicely provided expressions for these contributions in terms of complex integration contours or calculated them numerically under strong approximations, like low-frequency cut-offs \cite{Asenjo2017,Kornovan2016,Stourm2020,Sheremet2021}, but up to now how to compute these contributions exactly remains an open challenge.

In this paper, we address precisely this challenge by not only providing a comprehensive guide for the analytical calculation of guided and unguided mode contributions to the dipole-dipole interactions and collective dissipation induced among emitters trapped near a single mode optical nanofiber, but also supplying --to the best of our knowledge for the first time-- a method for the numerical calculation of the exact contribution of the radiation modes that overcomes convergence issues that typically make their accurate numerical determination extremely challenging. The method that we outline here can be generally used for the calculation of dipole-dipole interactions induced by other nanostructures, given that the mode profile functions are known \cite{PhysRevB.82.075427,Sipahigil2016,PhysRevA.95.033818,Albrecht2019,Vladimirova2021}. We show that, due to the presence of the nanofiber, the radiation field in its vicinity can be significantly altered, giving rise to contributions to the dipole-dipole interactions that significantly differ from the free-field counterparts. Owing to the collective character of both coherent and incoherent interactions, when measuring experimentally observable properties such as the transmission signal of fiber-guided light, these differences are even more evident as the number of emitters is increased.
\begin{figure}[t] 
  \begin{center}
      \includegraphics[width=\columnwidth]{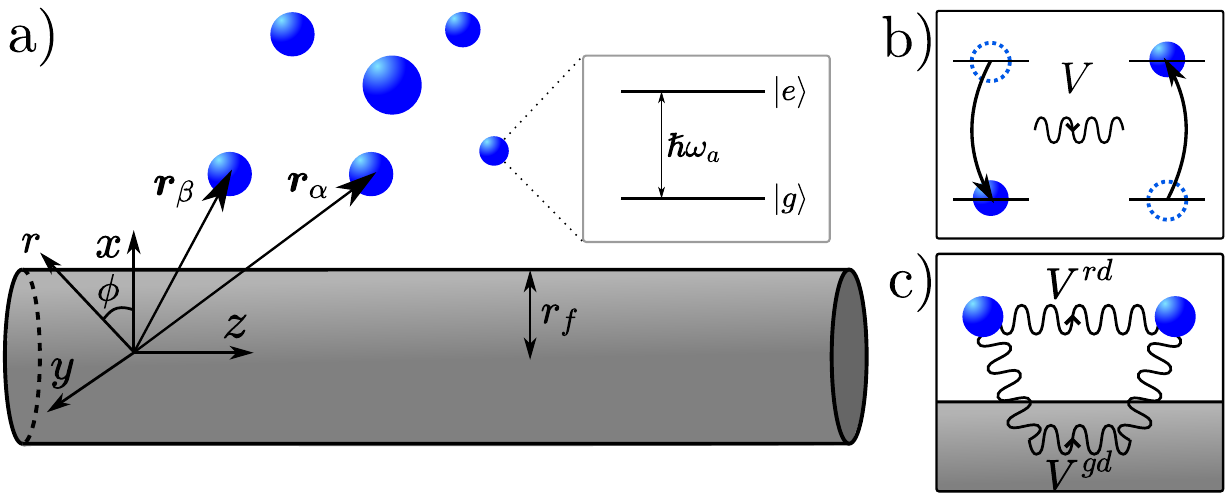}
  \end{center}
  \caption{\textbf{Dipole-dipole interactions between emitters next to a nanofiber.} a): $N$ emitters in vacuum are placed in the vicinity of a nanofiber with radius $r_f$. Each emitter is a two-level system with an excited ($|e\rangle$) and a ground ($|g\rangle$) state separated by an energy $\hbar \omega_a$. b): Due to their coupling to the electromagnetic field modes, the emitters interact through the exchange of virtual photons. c): These interactions have contributions stemming from the guided ($V^{gd}$) and the radiation ($V^{rd}$) modes.}
  \label{fig:NanofiberFig}
\end{figure}

\section{System and master equation}

We consider an ensemble of $N$ identical emitters placed near the surface of a nanophotonic waveguide (see Fig. \ref{fig:NanofiberFig}a). Each emitter is modelled as a two-level system with ground and excited states $\left|g\right>$ and $\left|e\right>$, respectively, separated by an energy $\hbar\omega_a$. Within the dipole and Born-Markov approximations, the dynamics of the reduced density matrix $\rho$, which contains the internal degrees of freedom of the emitters, is determined by the master equation \cite{PhysRevA.95.033818,PhysRevA.95.023838}
\begin{equation}\label{eq:meq}
\begin{split}
    \dot{\rho}=&i
    \sum_{\alpha\neq\beta}V_{\alpha \beta}[\sigma_\alpha^\dagger \sigma_\beta,\rho]\\
    &+\sum_{\alpha, \beta}\Gamma_{\alpha \beta}\left(\sigma_\beta\rho\sigma_\alpha^\dagger-\frac{1}{2}\{\sigma_\alpha^\dagger\sigma_\beta,\rho\}\right),
    \end{split}
\end{equation}
where $\sigma_\alpha = |g_\alpha\rangle \langle e_\alpha|$ is the spin-$1/2$ ladder operator for the $\alpha$-th emitter. The first term in the master equation describes dipole-dipole or exchange interactions, which are mediated by the exchange of virtual photons between the emitters (see Fig. \ref{fig:NanofiberFig}b). The exchange rate between the $\alpha$-th and $\beta$-th emitter is given by the magnitude of the coefficient $V_{\alpha \beta}$. The second term of (\ref{eq:meq}) describes incoherent photon emission or dissipation in the system, which in general possesses a collective character. Here, the diagonal terms of the dissipation matrix $\Gamma_{\alpha \beta}$ describe the single-emitter spontaneous decay rates, which for the $\alpha$-th emitter is thus given by $\Gamma_{\alpha \alpha}$. The sign and magnitude of the off-diagonal elements $\Gamma_{\alpha\beta}$ for $\alpha\neq\beta$ are responsible for the collective character of the emission, its landmark being the well-known super- and subradiant photon emission \cite{GROSS1982301,Crubellier_1985}.

Our aim is to obtain the coefficient matrices $V_{\alpha\beta}$ and $\Gamma_{\alpha\beta}$, which in turn fully determine all properties in the system. These can be determined via the electromagnetic Green's tensor $\bar{G}(\mathbf{r}_\alpha,\mathbf{r}_\beta,\omega)$ of the environment evaluated at the emitter positions $\mathbf{r}_\alpha$ and $\mathbf{r}_\beta$ as \cite{PhysRevA.95.033818,PhysRevB.82.075427,PhysRevA.66.063810}
\begin{equation*}
\begin{split}
    V_{\alpha \beta} &= \frac{1}{\pi\hbar\epsilon_0c^2}\mathcal{P}\!\int_{-\infty}^\infty\!\! d\omega\omega^2\frac{\mathbf{d}^*_\alpha\text{Im}\{\bar{G}(\mathbf{r}_\alpha,\mathbf{r}_\beta,\omega)\}\mathbf{d}_\beta^T}{\omega-\omega_a}, \\
    \Gamma_{\alpha \beta} &= \frac{2\omega^2_a}{\hbar\epsilon_0c^2}\mathbf{d}^*_\alpha\text{Im}\{\bar{G}(\mathbf{r}_\alpha,\mathbf{r}_\beta,\omega_a)\}\mathbf{d}^T_\beta,
\end{split}
\end{equation*}
where $\mathcal{P}$ denotes the principal value and $\mathbf{d}_\alpha$ is the transition dipole moment for the $\alpha$-th emitter. The electromagnetic Green's tensor is the solution to the wave equation
\begin{equation}\label{eq:waveeq}
    [k^2\epsilon(\mathbf{r},\omega)-\nabla\times\nabla\times]\bar{G}(\mathbf{r},\mathbf{r}',\omega)=
    -\delta(\mathbf{r}-\mathbf{r}')\mathbb{1},
\end{equation}
where $k=\omega/c$ and $\epsilon(\mathbf{r},\omega)$ is the dielectric constant of the medium. The Green's tensor satisfies the Schwarz reflection principle $\bar{G}^*(\mathbf{r},\mathbf{r}',\omega) = \bar{G}(\mathbf{r},\mathbf{r}',-\omega^*)$ and the Onsager reciprocity, $\bar{G}^T(\mathbf{r},\mathbf{r}',\omega) = \bar{G}(\mathbf{r}',\mathbf{r},\omega)$. As discussed in Appendix \ref{App:KK}, the complex function $\omega^2\bar{G}(\mathbf{r},\mathbf{r}',\omega)$ is by definition holomorphic in the complex upper half-plane and bounded at infinity. Hence, the dipole-dipole interaction may be written in terms of the real part of the Green's tensor
\begin{equation*}
    V_{\alpha \beta} = \frac{\omega_a^2}{\hbar\epsilon_0c^2}\mathbf{d}^*_\alpha\text{Re}\{\bar{G}(\mathbf{r}_\alpha,\mathbf{r}_\beta,\omega_a)\}\mathbf{d}^T_\beta,
\end{equation*}
where we have made use of the Kramers-Kronig relations. 

As shown above, for the calculation of the dipole-dipole and dissipation matrices we need to be able to separate the real and imaginary components of the Green's tensor, respectively. As the Green's tensors we consider are in general written as integrals over the frequency, to do so we introduce a small imaginary shift to the frequency $\bar{G}(\mathbf{r},\mathbf{r}',\omega) = \text{lim}_{\epsilon\rightarrow0^+} \bar{G}(\mathbf{r},\mathbf{r}',\omega+i\epsilon)$ \cite{LeKienGT} and use the Sokhotski–Plemelj theorem applied on the real line to find the imaginary part and compute the real part using the Kramers-Kronig relation. However, as discussed in Appendix \ref{App:KK}, this is valid only when the function we consider is well-defined and holomorphic in the frequency complex upper half-plane (including the real line). Decomposing the Green's tensor into its longitudinal and transverse components one can see that, while the transverse component is indeed well-behaved, the longitudinal component is singular at $\omega\to 0$ \cite{Knoell2001}. To illustrate this, let us consider the case where the medium is the vacuum (i.e. free space with no boundaries). Here, $\epsilon(\mathbf{r},\omega)=1$ everywhere and the solution to the wave equation \eqref{eq:waveeq} is known analytically and given by
\begin{eqnarray}\label{eq:vac}
        \bar{G}^0(\mathbf{r},\mathbf{r}',\omega) &=& \frac{ke^{i k\tilde{r}}}{4\pi}\biggl\{\frac{(\mathbb{1}-\hat{\tilde{\mathbf{r}}}^2)}{k\tilde{r}}\\ \nonumber
        &&+(3\hat{\tilde{\mathbf{r}}}^2-\mathbb{1}) \left[\frac{1}{(k\tilde{r})^3}-\frac{i}{(k\tilde{r})^2}\right]\biggl\} ,
\end{eqnarray}
where $\tilde{\mathbf{r}}=\mathbf{r}-\mathbf{r}'$, $\tilde{r} = |{\tilde{\mathbf{r}}}|$, and $\hat{\tilde{\mathbf{r}}} = {\tilde{\mathbf{r}}}/{\tilde{r}}$. Decomposing the vacuum Green's tensor into its longitudinal and transverse components as $\bar{G}^0(\omega)=\bar{G}^{0 \parallel}(\omega)+\bar{G}^{0 \perp}(\omega)$ \cite{Buhmann2012,Tiggelen2021} (see Appendix \ref{App:KK}), we observe that, while the transverse component is a holomorphic function in the complex upper half-plane, the longitudinal component is not, as it contains a second order pole at $\omega=0$. This in turn means that the usual form of the Kramers-Kronig relation cannot be used to find the real part of the full vacuum Green's tensor. It can rather only be validly applied to its tranverse part. With this, we find that
\begin{equation*}
    \text{Re}\{\bar{G}^{0}(\omega)\} = \frac{1}{\pi}\mathcal{P}\int_{-\infty}^{\infty}  d\omega'\frac{\text{Im}\{\bar{G}^0(\omega')\}}{\omega'-\omega}+\bar{G}^{0 \parallel}(\omega),
\end{equation*}
where we have used the fact that the longitudinal part is purely real. We will consider this relation when treating the more complicated situation of a cylindrical nanofiber.

\section{Green's tensor for a cylindrical nanofiber}

The Green's tensor can be expressed in terms of the eigenmodes of the positive frequency part of the electric field as
\begin{equation*}
    \bar{G}(\mathbf{r},\mathbf{r}',\omega) = \sum_n\frac{\mathbf{E}_n(\mathbf{r},\omega)\mathbf{E}^\dagger_n(\mathbf{r}',\omega)}{\lambda_n}.
\end{equation*}
Here, $\mathbf{E}_n(\mathbf{r},\omega)$ are the eigenvectors of the hermitian operator $\mathcal{H} = [k^2-\frac{1}{\epsilon(\mathbf{r},\omega)}\nabla\times\nabla\times]$ with eigenvalues $\lambda_n$. This representation allows us to easily split the contribution to the Green's tensor coming from the eigenmodes guided by the waveguide and the unguided (or radiative) ones as $\bar{G}(\mathbf{r},\mathbf{r}',\omega)=\bar{G}^{gd}(\mathbf{r},\mathbf{r}',\omega)+\bar{G}^{rd}(\mathbf{r},\mathbf{r}',\omega)$ \cite{BALIAN1970401,PhysRevA.43.467,PhysRevA.64.033812}, which leads in turn to the decomposition of the dipole-dipole and collective dissipation matrices: $V_{\alpha\beta}=V_{\alpha\beta}^{gd}+V_{\alpha\beta}^{rd}$ and $\Gamma_{\alpha\beta}=\Gamma_{\alpha\beta}^{gd}+\Gamma_{\alpha\beta}^{rd}$, see Fig. \ref{fig:NanofiberFig}c. Let us analyze these contributions separately for the case of a cylindrical dielectric nanofiber of radius $r_f$ characterized by a refractive index $n_1$ placed in an infinite vacuum of refractive index $n_2=1$. To do so, we will use a cylindrical coordinate system $(r,\phi,z)$, centered in the fiber core with the $z$-direction being along the fiber (see Fig. \ref{fig:NanofiberFig}a).

\subsection{Guided modes}

Using the expression for the guided modes of the electric field given in Appendix \ref{App:guided}, the guided contribution to the Green's tensor is given by 
\begin{equation*}
    \begin{split}
        \bar{G}^{gd}(\mathbf{r},\mathbf{r}',\omega')=&\frac{1}{2\pi}\sum_{fl} \int_0^{\infty}\!\! d\beta \frac{\mathbf{e}^{(\mu)}(\mathbf{r})\mathbf{e}^{(\mu)\dagger}(\mathbf{r}')}{{k'}^2-(\beta^2-q^2)} \\ &\times e^{il(\phi-\phi')}e^{if\beta(z-z')},
    \end{split}
\end{equation*}
where $(\mu)\equiv(\beta,l,f)$ are the labels of the guided modes. Here, $l=\pm 1$ is the polarization of the mode, $f=\pm 1$ is the propagation direction for the guided modes in the fiber and $\mathbf{e}^{(\mu)}(\mathbf{r})$ are the guided profile functions (note, that we assume that the nanofiber supports only the fundamental $\text{HE}_{11}$ guided mode). In addition, $\beta$ is the longitudinal propagation constant in the fiber, which for each value of $\omega$ is determined via the fiber eigenvalue equation (see Appendix \ref{App:guided}), $q$ is a variable associated to the field outside the nanofiber and $k' = \omega'/c$. The expression for the Green's tensor may be written in terms of the mode frequency $\omega$ by using the relation $q^2 = \beta^2-k^2$, such that
\begin{equation*}
\begin{split}
    \bar{G}^{gd}(\mathbf{r},\mathbf{r}',\omega')=&\frac{c^2}{2\pi}\sum_{fl} \int_0^\infty\!\! d\omega \frac{\mathbf{e}^{(\mu)}(\mathbf{r})\mathbf{e}^{(\mu)\dagger}(\mathbf{r}')}{{\omega'}^2-\omega^2}\\ &\times \beta'(\omega)e^{il(\phi-\phi')}e^{if\beta(\omega)(z-z')},
\end{split}
\end{equation*}
where $\beta'(\omega) = \frac{d \beta}{d\omega}$.

The guided contribution to the Green's tensor is purely transverse. We can therefore apply the Sokhotski–Plemelj theorem to find its imaginary part
\begin{equation*}
    \begin{split}
        \text{Im}\{\bar{G}^{gd}(\mathbf{r},\mathbf{r}',\omega)\} =& \frac{c^2}{4\omega}\sum_{fl} \mathbf{e}^{(\mu)}(\mathbf{r})\mathbf{e}^{(\mu)\dagger}(\mathbf{r}')\\
        &\times \beta'(\omega)e^{il(\phi-\phi')}e^{if\beta(\omega)(z-z')},
        \end{split}
\end{equation*}
which gives a guided contribution to the collective decay matrix that reads
\begin{equation*}
    \begin{split}
        \Gamma_{\alpha\beta}^{gd} =& \frac{\omega_a\beta_a'}{2\hbar\epsilon_0}\sum_{fl} \mathbf{d}^*_\alpha \cdot \mathbf{e}^{(\beta_a l f)}(\mathbf{r}_\alpha)\mathbf{d}_\beta \cdot \mathbf{e}^{(\beta_a l f)*}(\mathbf{r}_\beta)\\ &\times e^{il\phi_{\alpha\beta}}e^{if\beta_a z_{\alpha\beta}},
    \end{split}
\end{equation*}
with $\beta_a=\beta(\omega_a)$, $z_{\alpha\beta}=z_\alpha-z_\beta$ and $\phi_{\alpha\beta}=\phi_\alpha-\phi_\beta$.

The real part of the guided Green's tensor is found by using the usual Kramers-Kronig relation
\begin{equation*}
    \text{Re}\{\bar{G}^{gd}(\omega)\} =\frac{1}{\pi}\mathcal{P}\!\int_{-\infty}^\infty \!\!\!d\omega'\frac{\text{Im}\{\bar{G}^{gd}(\omega')\}}{\omega'-\omega}.
\end{equation*}
The principal value integral in this equation can be calculated analytically choosing an appropriate contour \cite{Solano,PhysRevA.95.023838}, and one finds that
\begin{equation*}
\begin{split}
    V_{\alpha\beta}^{gd} = &i\frac{\omega_a\beta_a'}{4\hbar\epsilon_0}\sum_{fl}\mathbf{d}_\alpha^*\cdot \mathbf{e}^{(\beta_a l f)}(\mathbf{r}_\alpha)\mathbf{d}_\beta\cdot \mathbf{e}^{(\beta_a l f)*}(\mathbf{r}_\beta)\\ &\times \text{sgn}(fz_{\alpha\beta})e^{il\phi_{\alpha\beta}}e^{if\beta_az_{\alpha\beta}},  
\end{split}
\end{equation*}
is the guided contribution to the dipole-dipole interaction.

\subsection{Radiation modes}

The radiation modes of the electric field can be both transverse and longitudinal, and the radiation contribution to the Green's tensor may be decomposed as $\bar{G}^{rd}(\omega) = \bar{G}^{rd \perp}(\omega)+\bar{G}^{rd \parallel}(\omega)$. On the other hand, one can also separate the Green's tensor as a sum of the vacuum modes and the ones scattered by the nanofiber as $\bar{G}^{rd}(\omega) = \bar{G}^{0}(\omega)+\bar{G}^{sc}(\omega)$. The scattered modes are, however, purely transverse, allowing to replace the longitudinal radiation Green's tensor with the well known longitudinal part of the vacuum Green's tensor and write
\begin{equation*}
    \bar{G}^{rd}(\mathbf{r},\mathbf{r}',\omega) = \bar{G}^{rd \perp}(\mathbf{r},\mathbf{r}',\omega)+\bar{G}^{0 \parallel}(\mathbf{r},\mathbf{r}',\omega).
\end{equation*}

Using the expression for the transverse radiation modes of the electric field given in Appendix \ref{App:unguided}, the radiative contribution to the Green's tensor is given by 
\begin{equation*}
    \begin{split}
        \bar{G}^{rd}(\mathbf{r},\mathbf{r}',\omega') &=  \sum_{ml}\int_{-\infty}^\infty \!\!d\beta \int_0^\infty \!\!dq \frac{\mathbf{e}^{(\nu)}(\mathbf{r})\mathbf{e}^{(\nu)\dagger}(\mathbf{r}')}{{k'}^2-(\beta^2+q^2)}\\ &\times e^{im(\phi-\phi')}e^{i\beta(z-z')}+\bar{G}^{0 \parallel}(\mathbf{r},\mathbf{r}',\omega'),
    \end{split}
\end{equation*}
where $(\nu) \equiv (\beta, q,m,l)$ are the labels of the radiation modes, with $l=\pm 1$ and $m=0,\pm 1, \pm 2,...$ labelling the polarization and order of the mode. In contrast to the guided mode case, $\beta = k\cos{\theta}$ is here a continuous variable for each value of the frequency $\omega$ and the variable $q = k\sin{\theta}$ is characteristic of the field outside the nanofiber. Finally, $\mathbf{e}^{(\nu)}(\mathbf{r})$ is the radiation profile function of the transverse electric field. The Green's tensor may be rewritten in terms of the mode frequency $\omega$ and the angle $\theta$ as
\begin{equation*}
\begin{split}
    \bar{G}^{rd}(\mathbf{r},&\mathbf{r}',\omega') = \sum_{ml}\int_0^\infty\!\! d\omega \omega\! \int_0^\pi\!\! d\theta \frac{\mathbf{e}^{(\nu)}(\mathbf{r})\mathbf{e}^{(\nu)\dagger}(\mathbf{r}')}{{\omega'}^2-\omega^2}\\
    &\times e^{im(\phi-\phi')}e^{i\frac{\omega}{c}\cos{\theta}(z-z')}+\bar{G}^{0 \parallel}(\mathbf{r},\mathbf{r}',\omega').
\end{split}
\end{equation*}
Note, that, as explained in Appendix \ref{App:unguided}, we have chosen specifically the eigenmode decomposition adopted in \cite{PhysRevA.64.033812}, which differs from, for example, the one adopted in \cite{PhysRevA.95.023838} in the accompanying normalization factors. However, let us point out that the two formulations are equivalent and hence give the same results.

Similarly as for the guided modes, we use the Sokhotski–Plemelj theorem to find the imaginary part of the radiation Green's tensor using the fact that the transverse component is holomorphic in the upper half complex plane and well-behaved on the real frequency axis. The imaginary part becomes then
\begin{equation*}
    \begin{split}
        \text{Im}\{\bar{G}^{rd}(\mathbf{r},\mathbf{r}',\omega)\} =& \frac{\pi}{2}\sum_{ml} \int_0^\pi d\theta \mathbf{e}^{(\nu)}(\mathbf{r})\mathbf{e}^{(\nu)\dagger}(\mathbf{r}')\\
        &\times e^{im(\phi-\phi')}e^{i\frac{\omega}{c}\cos{\theta}(z-z')},
    \end{split}
\end{equation*}
such that the corresponding contribution to the dissipation matrix reads
\begin{equation*}
    \begin{split}
        \Gamma^{rd}_{\alpha \beta} =& \frac{\pi \omega_a^2}{\hbar\epsilon_0c^2}\sum_{ml} \int_0^\pi d\theta \mathbf{d}_\alpha^* \cdot \mathbf{e}^{(\beta_a q_a m l)}(\mathbf{r}_\alpha)\\
        & \mathbf{d}_\beta \cdot \mathbf{e}^{(\beta_a q_a m l)*}(\mathbf{r}_\beta) e^{im\phi_{\alpha \beta}}e^{i\beta_a z_{\alpha \beta}},
    \end{split}
\end{equation*}
where now $\beta_a=\frac{\omega_a}{c}\cos{\theta}$ and $q_a=\frac{\omega_a}{c}\sin{\theta}$ are both functions of the angle $\theta$.

\begin{figure}[t] 
  \begin{center}
      \includegraphics[width=\columnwidth]{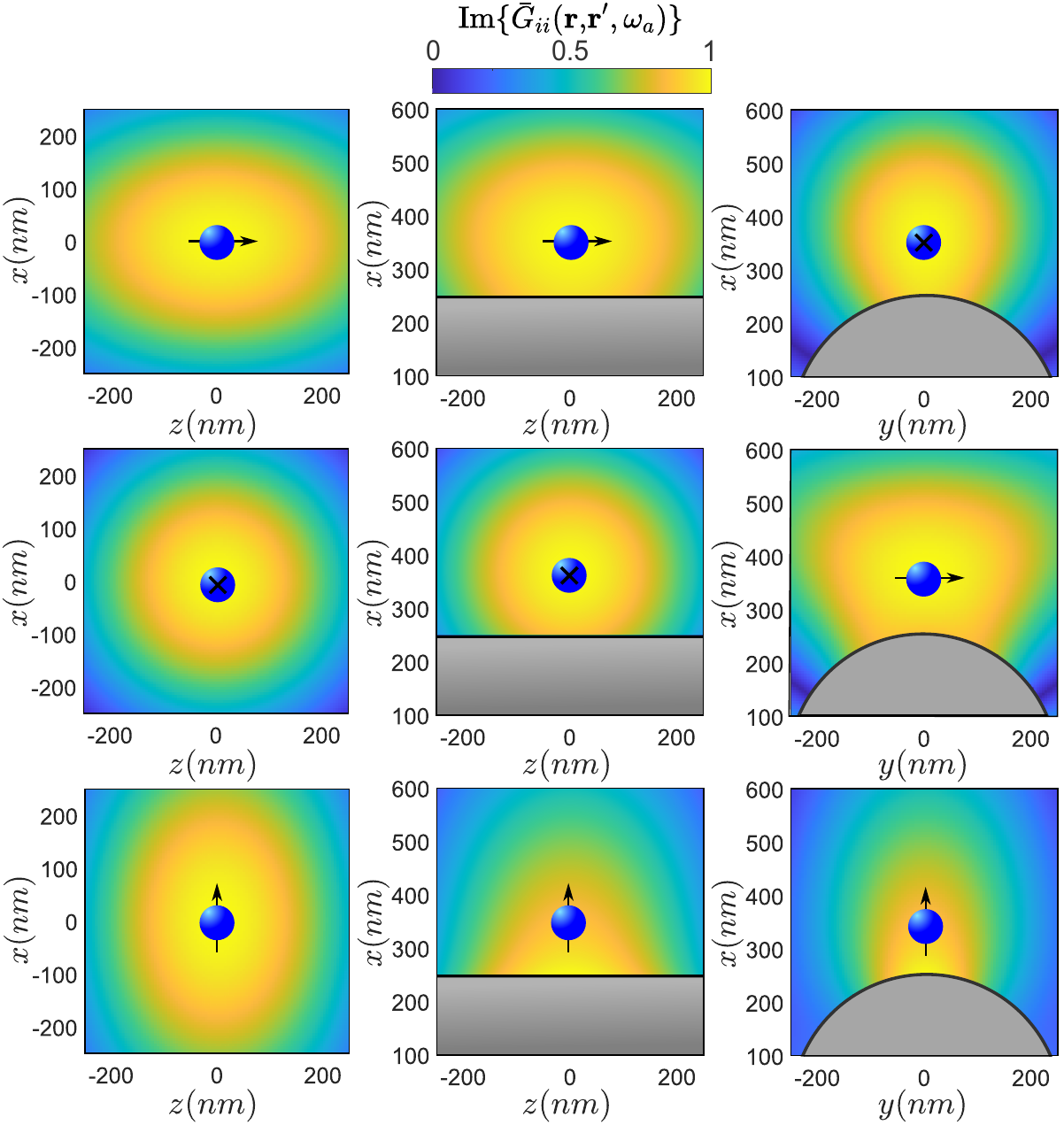}
  \end{center}
  \caption{\textbf{Modified radiation field near a nanofiber.} Diagonal component of the imaginary part of the Green's tensor $\text{Im}\{G_{ii}(\mathbf{r},\mathbf{r}',\omega_a)\}$ with $i=x,y,z$ (bottom, middle and upper row, respectively, all in arbitrary units) around a dipole placed at $\mathbf{r}'=0$ as a function of the coordinate $\mathbf{r}$ in the zx-plane (left and middle column) and in the yx-plane (right column). The cross in the middle of the dipole should be interpreted as the dipole pointing into the plane. The left column shows $\text{Im}\{G^0_{ii}\}$ in vacuum (free field), while the middle and right columns display $\text{Im}\{G^{rd}_{ii}\}$ at a distance of $100$ nm from the surface of a silica nanofiber with $r_f=250$ nm. The frequency $\omega_a=2\pi c/\lambda_a$ corresponds to the D2 transition in Cs, with $\lambda_a=852$ nm. The refractive index of the silica nanofiber is modeled using a Sellmeier equation \cite{Malitson:65}.}
  \label{ImG3x3}
\end{figure}

We use the imaginary part of the Green's tensor evaluated at $\omega=\omega_a$ to illustrate in Fig. \ref{ImG3x3} how much the presence of the nanofiber disturbs the field around a point dipole compared to the situation where the dipole is in free space. In particular, we can clearly observe that the field is most dramatically modified when the dipole moment is perpendicular to the axis of the nanofiber ($y$ and $x$ directions, second and third row in Fig. \ref{ImG3x3}).

The real part of the radiation Green's tensor can be found by applying the (modified) Kramers-Kronig relation to the transverse part of the radiation Green's tensor,
\begin{equation*}    
        \text{Re}\{\bar{G}^{rd}(\omega)\} \!=\! \frac{1}{\pi}\mathcal{P}\int_{-\infty}^{\infty}\!\!  d\omega'\frac{\text{Im}\{\bar{G}^{rd}(\omega')\}}{\omega'-\omega}+\bar{G}^{0 \parallel}(\omega),
\end{equation*}
such that the radiation contribution to the dipole-dipole interaction is given by
\begin{eqnarray}\nonumber
        V^{rd}_{\alpha \beta} &=& \frac{\omega_a^2}{\pi\hbar\epsilon_0 c^2}\mathcal{P}\int_{-\infty}^{\infty}  d\omega\frac{\mathbf{d}_\alpha^*\text{Im}\{\bar{G}^{rd}(\mathbf{r}_\alpha,\mathbf{r}_\beta,\omega)\}\mathbf{d}_\beta^T}{\omega-\omega_a}\\&&+\frac{\omega_a^2}{\hbar\epsilon_0 c^2}\mathbf{d}_\alpha^*\bar{G}^{0 \parallel}(\mathbf{r}_\alpha,\mathbf{r}_\beta,\omega_a)\mathbf{d}_\beta^T.
    \label{Vrd}
\end{eqnarray}
Due to the complicated frequency dependence of the imaginary part of the radiation Green's tensor, an analytical treatment of the principal value integral in equation \eqref{Vrd} is nontrivial and requires the introduction of branch cuts and the application of contour integration techniques \cite{Asenjo2017}. Due to these complications, in the literature typically this contribution is approximated by its free-field counterpart, which is known analytically \cite{Solano,PhysRevA.95.023838}. However, as illustrated by Fig. \ref{ImG3x3}, the field might actually be strongly modified by the presence of the nanofiber, and these modifications, even when small, may lead to large differences in observables when treating systems with many emitters. For these reasons, in the following we describe how to efficiently calculate numerically the exact value of $V^{rd}_{\alpha \beta}$. 

\subsection{Numerical evaluation of the radiative dipole-dipole interactions} 

\begin{figure}[t!] 
  \begin{center}
      \includegraphics[width=0.45\textwidth]{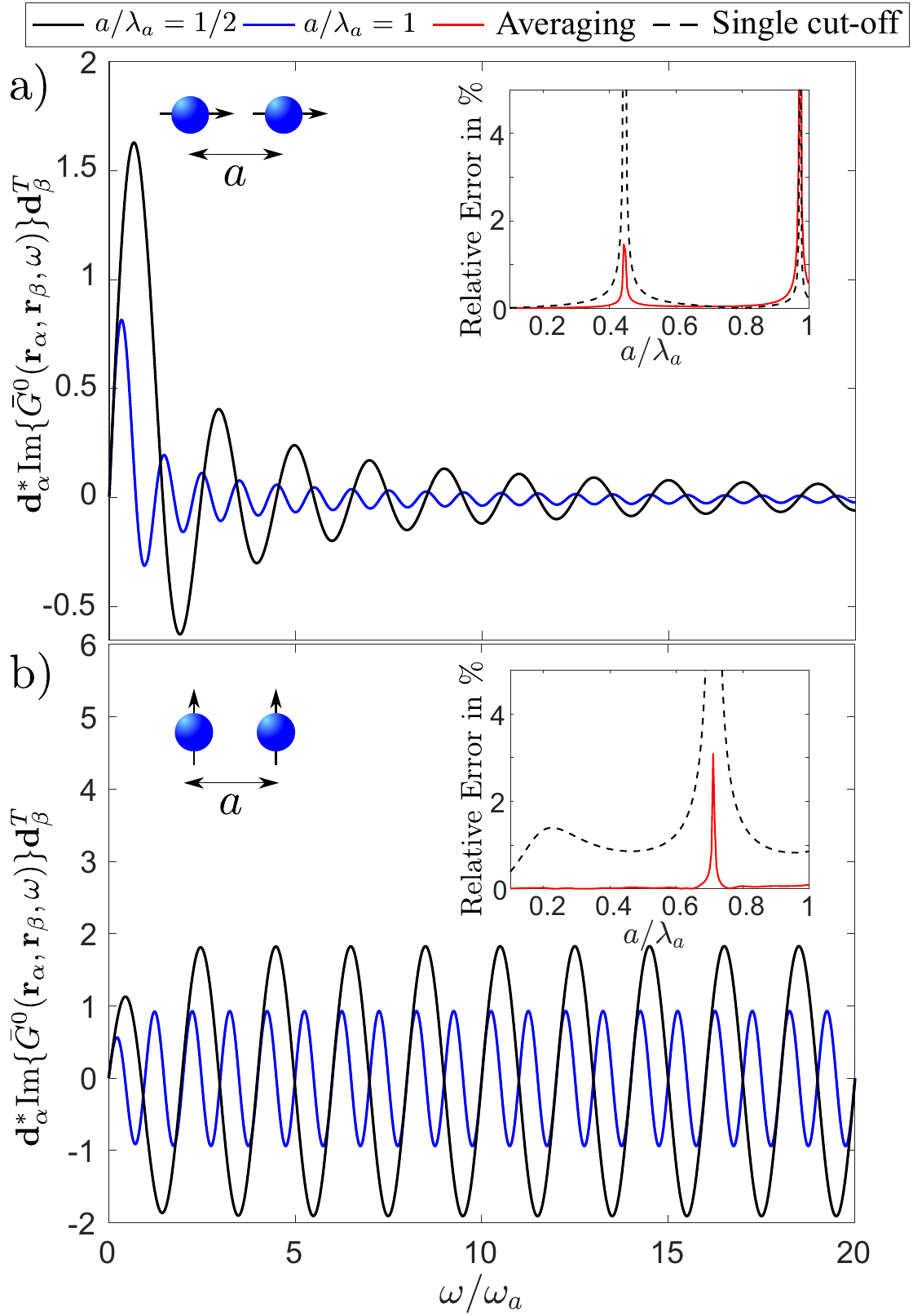}
  \end{center}
  \caption{\textbf{Numerical integration.} Imaginary part of the Green's tensor in vacuum (in arbitrary units) as a function of $\omega/\omega_a$ for two dipoles at a distance $a$ from one another pointing a) parallel and b) perpendicular to the displacement vector between them for $a/\lambda_a=1$ and $a/\lambda_a=1/2$ (black and blue solid lines, respectively). The insets show the relative error in $\%$ when calculating the vacuum dipole-dipole interaction $V_{\alpha\beta}^0$ as a function of $a/\lambda_a$ by using a direct numerical integration with a single cut-off frequency $\omega_c/\omega_a = 10\lambda_a/a$ (black dashed) and using the averaging method with a range of cut-offs $\omega_c/ \omega_a \in [2\lambda_a/a,10\lambda_a/a]$. The sharp peaks occur at the zeros of $V_{\alpha\beta}^0$.}
  \label{Error}
\end{figure}

The first challenge one encounters to evaluate numerically the integral in (\ref{Vrd}) is the singularity in the integrand, happening at $\omega=\omega_a$. This singularity can be avoided by using a fast Fourier transform, as introduced in Refs. \cite{Peterson:73,Saxton_1974}, and using the Schwarz reflection principle, such that $V^{rd}_{\alpha \beta}$ may be written as
\begin{eqnarray}\nonumber
    V^{rd}_{\alpha \beta} &=& \frac{ 2\omega_a^2}{\pi\hbar\epsilon_0c^2}\bigg[\int_{0}^{\infty}\!\!d\omega \mathbf{d}^*_\alpha\text{Im}\{\bar{G}^{rd}(\mathbf{r}_\alpha,\mathbf{r}_\beta,\omega)\}\mathbf{d}_\beta^T\\\label{VwithFFT}
    &&\times\int_0^{\infty}\!\!d\tau \cos{(\omega_a\tau)}\sin{(\omega\tau)}\\&&+\frac{\pi}{2}\mathbf{d}_\alpha^*\bar{G}^{0 \parallel}(\mathbf{r}_\alpha,\mathbf{r}_\beta,\omega_a)\mathbf{d}_\beta^T\bigg].\nonumber
\end{eqnarray}
However, the main obstacle for the calculation of $V_{\alpha\beta}^{rd}$ lies in the imaginary part of the Green's tensor, in particular in its behavior at large values of $\omega$. To illustrate this, let us consider again the Green's tensor in vacuum (\ref{eq:vac}). In the limit of large $\omega$, i.e. when $k\tilde{r}\gg 1$, the imaginary part of this tensor yields
\begin{equation*}
\begin{split}
    \text{Im}\{\bar{G}^{0}(\mathbf{r},\mathbf{r}',\omega)\} \approx& \frac{1}{4\pi\tilde{r}}\!\bigg[(\mathbb{1}-3\hat{\tilde{\mathbf{r}}}^2) \frac{\cos{k\tilde{r}}}{k\tilde{r}}\\
    &+(\mathbb{1}-\hat{\tilde{\mathbf{r}}}^2)\sin{k\tilde{r}}\bigg].
    \end{split}
\end{equation*}
Here we can see that the integrand in the frequency integral in (\ref{VwithFFT}) for large $\omega\gg\omega_a$ is an oscillating function with period proportional to $\omega_a \lambda_a/r_{\alpha\beta}$, where $\lambda_a=\frac{2\pi c}{\omega_a}$ is the transition wavelength. The amplitude of the oscillations either decays very slowly, as $1/\omega$ (e.g. in vacuum when the dipoles are aligned with the displacement vector, i.e. $|\hat{\mathbf{d}}_{\alpha}\cdot\hat{\tilde{\mathbf{r}}}_{\alpha \beta}|=1$, see Fig. \ref{Error}a), or is constant. In the former case, reaching convergence of the integral over frequency requires then an extremely high cut-off frequency $\omega_c\gg\omega_a \lambda_a/r_{\alpha\beta}$, which grows the closer the emitters are to each other. Moreover, in general, even a large value for $\omega_c$ does not necessarily ensure convergence (see e.g. Fig. \ref{Error}b for emitters in the free-field with $|\hat{\mathbf{d}}_{\alpha}\cdot\hat{\tilde{\mathbf{r}}}_{\alpha \beta}|=0$, where the amplitude of the oscillations remains constant for large $\omega$).

In addition to this issue, note that for the numerical calculation of the imaginary part of the Green's tensor in the presence of the cylindrical nanofiber, in principle an infinite sum of modes $m$ should be considered. In practice, the sum is truncated at some finite value $m_c$. However, for a fixed distance of the emitters with respect to the nanofiber surface, the amount of modes necessary to ensure convergence grows dramatically with increasing $\omega$. This means that the computational cost of increasing the cut-off frequency is enormous. In order to circumvent this problem, we make use of the periodic nature of the integrand in the limit of large $\omega$. This allows us to approximate the result of the frequency integral as the average of the outcomes for a range of cut-off frequencies comprising a few oscillations. This results in a much faster convergence of the integral when the integrand decays as $1/\omega$, as the maximum cut-off frequency necessary for convergence is relatively low. Most importantly, when the amplitude of the oscillations remains constant, this allows to obtain convergence which would remain otherwise elusive.

In order to illustrate the power of this approach, we have applied it to calculate numerically the dipole-dipole interactions in vacuum. We show in the insets of Fig. \ref{Error} the relative error of the numerical calculation compared to the exact one (known analytically in this case), using a "single cut-off" direct integration of \eqref{Vrd} and an "averaging" method for the frequency integration in expression (\ref{VwithFFT}). It is evident that the second approach gives dramatically better results, particularly when the dipoles are perpendicular to their displacement vector. This encourages us to apply this numerical method to the nanofiber case, where the main features of the Green's tensor as a function of the frequency are similar to the vacuum ones.

We have made publicly available a code that follows this recipe to calculate both guided and radiation contributions to both dipole-dipole and incoherent interactions in \cite{code}.

\subsection{Modified dipole-dipole interaction between two atoms near a nanofiber}

\begin{figure}[t]
  \begin{center}
      \includegraphics[width=0.40\textwidth]{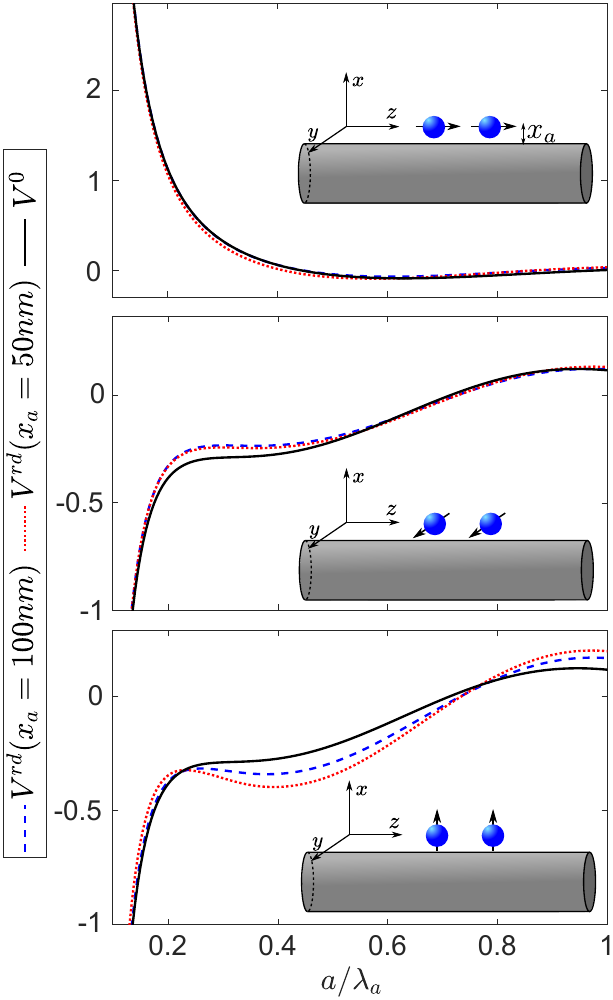}
  \end{center}
  \caption{\textbf{Modified dipole-dipole interactions.} The radiation contribution to the dipole-dipole interaction $V^{rd}$ (in units of the single particle decay rate $\gamma$) for a distance of $x_a = 50$ nm (red dotted) and $x_a=100$ nm (blue dashed) from the fiber surface as a function $a/\lambda_a$. The black solid line is the analytic solution in the case of vacuum. Each panel displays the results for a different orientation of the dipoles with respect to the surface of the nanofiber: parallel, binormal and normal for the upper, middle and lower panel, respectively. The parameters are identical to those used in Fig. \ref{ImG3x3}.}
  \label{VrdFigComb}
\end{figure}

We apply here the numerical method presented above to the simple case of two identical two-level atoms placed at a fixed distance from the fiber surface $x_a$, fixed azimuthal angle $\phi=0$ and fixed separation $a$. Figure \ref{VrdFigComb} shows the radiation contribution to the dipole-dipole interaction $V^{rd}$ for distances $x_a=50$nm and $x_a=100$nm from the fiber surface compared to the vacuum counterpart $V^0$ as a function of $a/\lambda_a$ for three different dipole orientations. The dipole-dipole interaction in all cases is normalized by the single-atom spontaneous decay rate in vacuum, $\gamma = |\mathbf{d}|^2\omega_a^3/(3c^3\pi \epsilon_0 \hbar)$.

As expected from the results for the imaginary part of the Green's tensor (Fig. \ref{ImG3x3}), the most important modifications of the dipole-dipole interactions are found when the atomic dipole moment orientation is perpendicular to both the axis and the surface of the nanofiber (bottom panel). Here, the differences between the calculated radiation components and their vacuum counterparts are not only large (e.g., 70\% difference at $a/\lambda_a=1$), but also persist at large distances between the atoms i.e. at $a\geq\lambda_a$.

Finally, to further illustrate the power of our approach, let us consider a pair of dipoles separated by $a=\lambda$ parallel to the displacement vector (Figs. \ref{Error}a and \ref{VrdFigComb}a). Using our numerical method, here we calculate the vacuum interaction to a precision of 0.004$\%$ error using an upper cut-off frequency of $\omega_c = 10\omega_a$. To get the same precision using a standard numerical integration method, a cut-off frequency of $\omega_c = 40\omega_a$ would be needed. For $\omega = 10\omega_a$ we have to include around $m_c=90$ modes, while for $\omega = 40\omega_a$ around $260$ modes are needed (needing about six times longer to calculate the imaginary part of the Green's tensor at that frequency point alone).

\section{Effect of the collective interaction on the transmission of fiber-guided light}

The effect of the modifications that we find in the dipole-dipole interaction between a pair atoms becomes greatly amplified when considering observables in a large ensemble of many emitters. The key to understand this amplification is that the emitters couple \textit{collectively} to its environment. This in turn means that each of the $N^2$ terms $V_{\alpha\beta}$ and $\Gamma_{\alpha\beta}$ of the dipole-dipole and dissipation matrices play a role in the (collective) coherent and incoherent dynamics of the system. Hence, small variations of the coefficients $V_{\alpha\beta}$ and $\Gamma_{\alpha\beta}$ in the master equation (\ref{eq:meq}) have a large effect on observables such as the photon emission rate or the spectrum of the light that is absorbed and emitted from the ensemble.
\begin{figure*}[t!] 
  \begin{center}
      \includegraphics[width=0.9\textwidth]{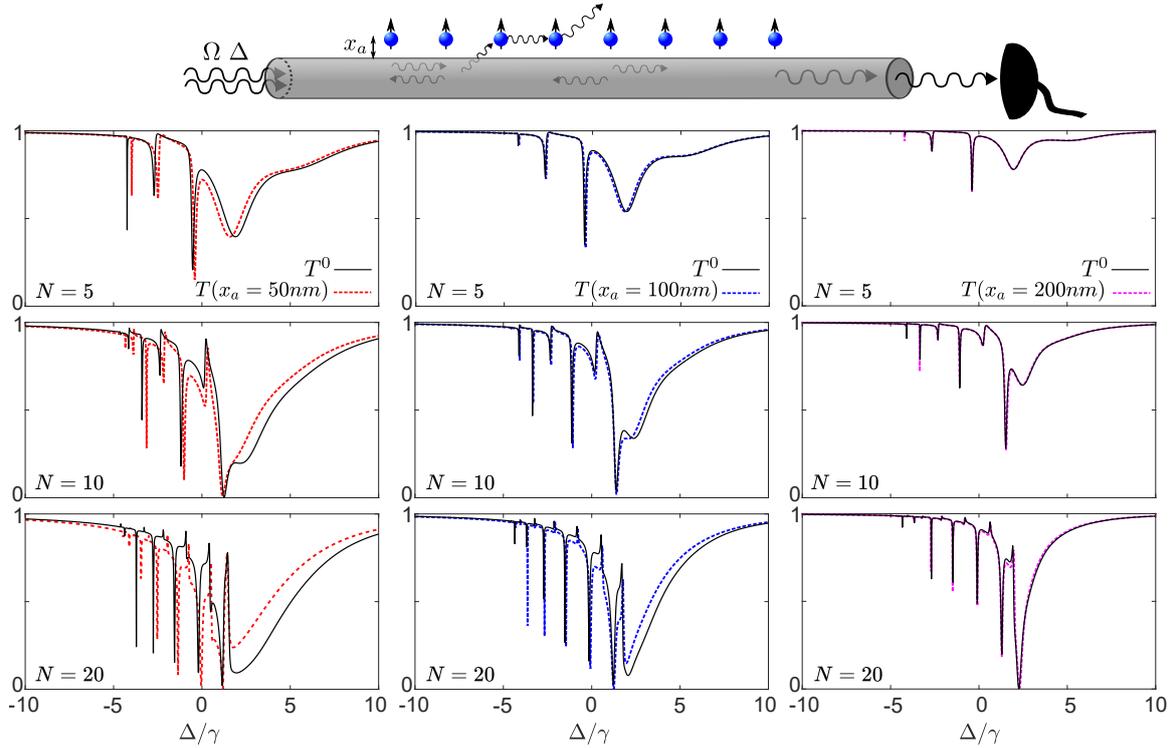}
  \end{center}
  \caption{\textbf{Modified transmission signal.} Transmission signal $T$ in the stationary state of fiber-guided light driving the atoms with Rabi frequency $\Omega$ and detuning $\Delta$ for a chain of $N$ atoms with lattice constant $a = 0.1\lambda_a$ placed at a distance $x_a$ from the surface of the nanofiber. The black solid line represents the transmission $T^0$ using the vacuum instead of the full radiation dipole-dipole interactions.}
  \label{Transmission3x3}
\end{figure*}

We illustrate this now by investigating the transmission signal of fiber-guided light when a periodic chain of $N$ atoms is placed in the vicinity of the fiber surface (see Fig. \ref{Transmission3x3}). Here, in the cylindrical coordinate system, the coordinates of atom $\alpha$ are given by $\mathbf{r}_{\alpha}=(x_a,0,(\alpha-1)a)$. The system is driven by a weak probing field of frequency $\omega_p$ detuned from the atomic resonance frequency by the detuning $\Delta = \omega_p-\omega_a$ and with Rabi frequency $\Omega$. The light field enters the nanofiber from its left, is guided through the nanofiber, interacts with the atomic chain, and its transmitted signal is measured at a position $z$ to the right of the chain in the fiber as \cite{PhysRevA.95.033818,Asenjo2017} 
\begin{equation}\label{eq:T}
    T = \frac{\langle \psi | E^{gd \dagger}_R(z) E^{gd}_R(z) | \psi \rangle}{\Omega^2}.
\end{equation}
Here $\hat{E}^{gd}_R$ is the right propagating component of the guided probe field, given by a sum of the input and scattering components as
\begin{equation*}
    E^{gd}_R(z) = \Omega e^{i\beta z}+i\frac{\gamma^{gd}}{2}\sum_{\alpha=1}^{N}\Theta(z-z_\alpha)e^{i\beta(z-z_\alpha)}\sigma_{\alpha},
\end{equation*}
where $\gamma^{gd} = \Gamma^{gd}_{\alpha\alpha}$ is the single-atom decay rate of each atom into the guided modes.

In the single excitation sector, the atomic wave function may be written as $|\psi(t) \rangle =c_G(t)|G \rangle +\sum_{\alpha=1}^N c_e^\alpha(t)|e_\alpha \rangle$, where $\left|G\right>=\left|g_1\right>\otimes\left|g_2\right>\dots\left|g_N\right>$ and $\left|e_\alpha\right>=\left|g_1\right>\dots\otimes\left|e_\alpha\right>\otimes\dots\left|g_N\right>$. In the low saturation regime (weak probe laser), $c_G(t)\approx 1$, and inserting the ansatz into the master equation (\ref{eq:meq}), the time evolution of the probability amplitude coefficients $c_e^\alpha(t)$ can be found to be determined by
\begin{equation*}    
    \dot{\boldsymbol{c}}_e(t) = i\left(\Delta+i\frac{\Gamma}{2}-H^\mathrm{eff}\right)\boldsymbol{c}_e(t)+i\boldsymbol{\eta},
\end{equation*}
where the components of $\boldsymbol{\eta}$ are $\eta_\alpha=\Omega e^{i\beta z_\alpha}$ and $\Gamma=\Gamma_{\alpha\alpha}$ is the total single-atom decay rate (sum of the decay rates into guided and radiation modes). Finally, the effective Hamiltonian $H^\mathrm{eff}$ given by
\begin{equation*}
    H^{\text{eff}}_{\alpha\beta} = -V_{\alpha\beta}-i\frac{\Gamma_{\alpha\beta}}{2},
\end{equation*}
has been introduced for $\alpha\neq\beta$.

In Figure \ref{Transmission3x3}, we show the guided light transmission signal (\ref{eq:T}) as a function of $\Delta$ in the stationary state [where $\dot{\boldsymbol{c}}_e(t)=0$] when a periodic chain of $N=5,10$ and 20 atoms with nearest neighbor separation $a/\lambda_a=0.1$ is placed at $x_a=50, 100$ and 200 nm from the surface of the nanofiber. We compare in all cases the transmission with the result obtained under the approximation that the radiation component of the dipole-dipole interaction is given by the free-field $V^0_{\alpha\beta}$. One can easily observe here that indeed large deviations arise in the transmission spectrum. These differences become more evident the closer the atoms are to the nanofiber (where each element $V^{rd}_{\alpha\beta}$ differs more from $V^{0}_{\alpha\beta}$, see Fig. \ref{VrdFigComb}) and the larger the number $N$ of atoms in the chain (where more elements $V^{rd}_{\alpha\beta}$ participate in the collective dynamics). In all cases one can observe, not only measurable shifts of the (subradiant) narrow and (superradiant) broad resonance peaks, but an overall deformation of the spectrum as $N$ increases. Note, that in current experiments typically several hundreds of atoms can be trapped near the nanofiber \cite{Liedl2022}, making the spectra differ even more substantially.

\section{Conclusions and outlook}

We have provided a detailed recipe for the analytical and numerical calculation of the interaction of an ensemble of emitters in the presence of a cylindrical nanofiber. In particular, we have shown that the dipole-dipole interactions mediated by the radiation modes outside the nanofiber differ significantly (up to a 70\% in some of the cases considered) from the ones obtained in vacuum, specially when the emitters' transition dipole moments are oriented normal or binormal to the fiber surface. These differences affect substantially the collective properties of the system, such as the transmission signal of fiber-guided light, where large shifts in the resonances are observed as the number of emitters is increased. Note, however, that we have calculated the transmission in the single-excitation limit (linear optics regime). Beyond this limit, we expect that, while features related to superradiance will not dramatically affected by modifications in the dipole-dipole interactions \cite{Masson2022}, these will become crucial when studying the excitation and description of many-body long-lived (subradiant) states, e.g. for the realisation of quantum memories \cite{Sayrin2015,Goraud2015}.

\begin{acknowledgments}
The authors thank I. Lesanovsky for insightful comments and discussions. The authors acknowledge support by the state of Baden-Württemberg through bwHPC and the German Research Foundation (DFG) through grant no INST 40/575-1 FUGG (JUSTUS 2 cluster). We acknowledge support by Open Access Publishing Fund of University of Tübingen. The research leading to these results has received funding from the Deutsche Forschungsgemeinsschaft (DFG, German Research Foundation) under Project No. 452935230 and the search Unit FOR 5413/1, Grant No. 465199066.
\end{acknowledgments}

\appendix

\section{The Green's tensor and the applicability of the Kramers-Kronig relation}\label{App:KK}

In this Appendix we discuss the applicability of the Kramers-Kronig relation for the Green's tensor.

The application of the Kramers-Kronig relation to a complex function $\chi (\omega)$ requires that the function satisfies two conditions. The function has to be holomorphic in the upper half plane, i.e. the function contains no poles in the complex upper half plane and the function has to converge to a finite value as $|\omega| \rightarrow \infty$ in the upper half plane. 

The Green's tensor converges as $|\omega| \rightarrow \infty$ in the upper half plane, but it can have poles on the real line at $\omega = 0$. Hence, the Green's tensor does not in general satisfy the conditions for the Kramers-Kronig relation. However by multiplying the Green's tensor by $\omega^2$, the small and large frequency limits may be applied \cite{PhysRevA.68.043816}
\begin{equation*}
\begin{split}
    \lim_{|\omega| \to 0} \frac{\omega^2}{c^2}\bar{G}(\mathbf{r},\mathbf{r}',\omega) &= -\mathbb{1}\delta(\mathbf{r}-\mathbf{r}'), \\ \lim_{|\omega| \to \infty} \frac{\omega^2}{c^2}\bar{G}(\mathbf{r},\mathbf{r}',\omega) &= \bar{M},
\end{split}
\end{equation*}
where the components of $\bar{M}$ are $M_{ij} < \infty$, which is exactly the Kramers-Kronig conditions. Therefore, the Kramers-Kronig relation may always be applied to the function  $\omega^2\bar{G}(\mathbf{r},\mathbf{r}',\omega)$, but not always to the Green's tensor itself. 

However, the singular part of the Green's tensor at $\omega = 0$ may be isolated by decomposing the Green's tensor into its transverse and longitudinal components
\begin{equation*}
    \bar{G}(\mathbf{r},\mathbf{r}',\omega) = \bar{G}^{\perp}(\mathbf{r},\mathbf{r}',\omega) + \bar{G}^{\parallel}(\mathbf{r},\mathbf{r}',\omega).
\end{equation*}
The transverse Green's tensor is in general holomorphic in the complex upper half-plane including the real line and hence satisfies the conditions for the Kramers-Kronig relation, while the longitudinal part is singular in $\omega = 0$. The transverse and longitudinal part of the Green's tensor may be found by means of the transverse and longitudinal projection operators \cite{Knoell2001,NICOROVICI20102915}
\begin{equation*}
\begin{split}
    \bar{G}^{\perp}(\mathbf{k},\omega) &= \bar{\Pi}^{\perp}\bar{G}(\mathbf{k},\omega)\bar{\Pi}^{\perp}, \\
    \bar{G}^{\parallel}(\mathbf{k},\omega) &= \bar{\Pi}^{\parallel}\bar{G}(\mathbf{k},\omega)\bar{\Pi}^{\parallel},
\end{split}
\end{equation*}
where the projection operators are defined as 
\begin{equation*}
    \bar{\Pi}^{\perp}_{ij} = \delta_{ij}- \frac{k_ik_j}{|\mathbf{k}|^2}, \qquad     \bar{\Pi}^{\parallel}_{ij} = \frac{k_ik_j}{|\mathbf{k}|^2}.
\end{equation*}
As an example, let us now decompose the vacuum Green's tensor into its longitudinal and transverse components as \cite{Buhmann2012,Tiggelen2021}
\begin{align*}
\begin{split}
    \bar{G}^{0 \parallel}(\tilde{\mathbf{r}},\omega) =& (3\hat{\tilde{\mathbf{r}}}^2-\mathbb{1})\frac{1}{4\pi \tilde{r}^3k^2},
\end{split}
\\
\begin{split}
    \bar{G}^{0 \perp}(\tilde{\mathbf{r}},\omega) =& \biggl\{(3\hat{\tilde{\mathbf{r}}}^2-\mathbb{1}) \left[\frac{1}{(k\tilde{r})^3}-\frac{i}{(k\tilde{r})^2}\right]\\&+(\mathbb{1}-\hat{\tilde{\mathbf{r}}}^2)\frac{1}{k\tilde{r}}\biggl\}\frac{ke^{ik\tilde{r}}}{4\pi} \\ &-(3\hat{\tilde{\mathbf{r}}}^2-\mathbb{1})\frac{1}{4\pi \tilde{r}^3k^2},
\end{split}
\end{align*}
where $\tilde{\mathbf{r}}=\mathbf{r}-\mathbf{r}'$, $\tilde{r} = |{\tilde{\mathbf{r}}}|$ and $\hat{\tilde{\mathbf{r}}} = {\tilde{\mathbf{r}}}/{\tilde{r}}$. Here, we can see that, while the transverse component of the Green's tensor is a holomorphic function in the complex upper half-plane, the longitudinal component is not, as it contains a second order pole at $\omega=0$.

\section{Guided modes}\label{App:guided}
The guided modes of the nanofiber are denoted by $(\mu) = (\beta,l,f)$, where $\beta$ is the longitudinal propagation constant, $l=\pm 1$ is the polarization and $f=\pm 1$ is the propagation direction in the fiber. The guided modes of the electric field at a position $\mathbf{r} = (r,\phi,z)$ may be expressed as
\begin{equation*}
\mathbf{E}^{(\mu)}(\mathbf{r}) = \mathbf{e}^{(\mu)}(r)e^{il\phi}e^{if\beta z},
\end{equation*}
where $\mathbf{e}^{(\mu)}(r)$ is the guided profile function of the fiber with components in cylindrical coordinates, for $r<r_f$, given by
\begin{equation*}
    \begin{split}
        e^{(\mu)}_r(r) =& iC\frac{q}{\kappa}\frac{K_1(qr_f)}{J_1(qr_f)}[(1-s)J_0(\kappa r)\\&-(1+s)J_2(\kappa r)], \\
        e^{(\mu)}_\phi(r) =& -C\frac{lq}{\kappa}\frac{K_1(qr_f)}{J_1(qr_f)}[(1-s)J_0(\kappa r)\\&+(1+s)J_2(\kappa r)], \\
        e^{(\mu)}_z(r) =& C\frac{2fq}{\kappa}\frac{K_1(qr_f)}{J_1(qr_f)}J_1(\kappa r),
    \end{split}
\end{equation*}
and for $r>r_f$
\begin{equation*}
    \begin{split}
        e^{(\mu)}_r(r) =& iC[(1-s)K_0(qr)+(1+s)K_2(qr)], \\
        e^{(\mu)}_\phi(r) =& Cl[(1+s)K_2(qr)-(1-s)K_0(qr)], \\
        e^{(\mu)}_z(r) =& C\frac{2fq}{\beta}K_1(qr).
    \end{split}
\end{equation*}
Here, $\kappa^2 = k^2n_1^2-\beta^2$ and $q^2 =\beta^2-k^2$ characterizes the field inside and outside the fiber respectively, $k=\omega/c$ is the free space propagation constant and $n_{1}$ is the refractive index of the fiber. The longitudinal propagation constant $\beta$ of the fiber may be determined by solving the fiber eigenvalue equation
\begin{equation*}
    \begin{split}
    \frac{J_0(\kappa r_f)}{\kappa r_f J_1(\kappa r_f)} &= \frac{n_1^2+1}{2n_1^2} \frac{K'_1(q r_f)}{q r_f K_1(q r_f)}+ \frac{1}{\kappa^2r_f^2} \\ &-\biggl[\left(\frac{n_1^2-1}{2n_1^2}\frac{K'_1(q r_f)}{q r_f K_1(q r_f)}\right)^2\\&+\frac{\beta^2}{n_1^2k^2}\left(\frac{1}{q^2r_f^2}+\frac{1}{\kappa^2r_f^2}\right)^2 \biggl]^{1/2}.
\end{split}
\end{equation*}
The functions $J_m(x)$ and $K_m(x)$ are the Bessel function of first kind and modified Bessel function of second kind respectively, which has the following form in the asymptotic limit ($|x| \gg 1$)
\begin{equation*}
    \begin{split}
    J_m(x) &\approx \sqrt{\frac{2}{\pi x}}\cos{\left(x-\frac{m\pi}{2}-\frac{\pi}{4}\right)}, \\
    K_m(x) &\approx \sqrt{\frac{2}{\pi x}}e^{-x}.
    \end{split}
\end{equation*}
The primed Bessel functions are their derivatives $J'_m(x)=\frac{d}{dx}J_m(x)$, $K'_m(x)=\frac{d}{dx}K_m(x)$. The parameter $s$ is defined as
\begin{equation*}
    s = \frac{\frac{1}{\kappa^2r_f^2}+\frac{1}{q^2r_f^2}}{\frac{J'_1(\kappa r_f)}{\kappa r_f J_1(\kappa r_f)}+ \frac{K'_1(q r_f)}{q r_f K_1(q r_f)}}.
\end{equation*}

$\mathbf{E}^{(\mu)}(\mathbf{r})$ are eigenvectors of the Hermitian operator $\mathcal{H} = [k^2-\frac{1}{\epsilon(\mathbf{r},\omega)}\nabla\times\nabla\times]$ with eigenvalues $\lambda_{(\mu)} = k_0^2-(\beta^2-q^2)$, where $k_0 = \omega_0/c$ and $\omega_0$ is the argument frequency of the guided Green's tensor.

The profile function of a guided mode is
\begin{equation*}
    \mathbf{e}^{(\mu)}(r) = e^{(\mu)}_r(r)\hat{\mathbf{r}}+e^{(\mu)}_\phi(r)\hat{\boldsymbol{\phi}}+e^{(\mu)}_z(r)\hat{\mathbf{z}},
\end{equation*}
and is normalized
\begin{equation*}
    \int_0^{2\pi}d\phi\int_0^{\infty}\epsilon(\mathbf{r},\omega)|\mathbf{e}^{(\mu)}|^2rdr=1.
\end{equation*}

The components of the guided profile function has the following symmetry relations
\begin{equation*}
    \begin{split}
        e^{(\beta l f)}_r &= e^{(\beta l -f)}_r = e^{(\beta -l f)}_r, \\
        e^{(\beta l f)}_\phi &= e^{(\beta l -f)}_\phi = -e^{(\beta -l f)}_\phi, \\
        e^{(\beta l f)}_z &= -e^{(\beta l -f)}_z = e^{(\beta -l f)}_z.
    \end{split}
\end{equation*}

\section{Radiation modes}\label{App:unguided}

The radiation modes of the nanofiber are denoted by $(\nu) = (\beta,q,m,l)$, where $\beta$ is the longitudinal propagation constant, $m = 0,\pm 1, \pm 2,...$ is the mode order and $l = \pm 1$ is the polarization. The characteristics of the field inside and outside the fiber are $\kappa^2 = k^2n_1^2-\beta^2$ and $q^2=k^2-\beta^2$ respectively. Unlike for the guided modes, the variables $\beta$ and $q$ for the radiation modes are continuous variables for each value of $k$ and may be expressed in terms of an angle $\theta$
\begin{equation*}
    \beta=k\cos{\theta}, \qquad q=k\sin{\theta}.
\end{equation*}
The radiation modes of the electric field at position $\mathbf{r} = (r,\phi,z)$ may be expressed as 
\begin{equation*}
    \mathbf{E}^{(\nu)}(\mathbf{r}) = \mathbf{e}^{(\nu)}(r)e^{im\phi}e^{i\beta z}.
\end{equation*}
We will only consider the transverse modes of the radiation electric field, $\nabla\cdot[\epsilon(\mathbf{r})\mathbf{E}^{(\nu)}(\mathbf{r})]=0$. In this case, $\mathbf{e}^{(\nu)}(r)$ is the transverse radiation profile function of the fiber with components in cylindrical coordinates, for $r<r_f$, given by
\begin{equation*}
    \begin{split}
        e_r^{(\nu)}(r) =& \frac{i\beta}{\kappa} A J'_m(\kappa r) -m\frac{\omega \mu_0}{\kappa^2r}BJ_m(\kappa r), \\
        e_\phi^{(\nu)}(r) =& - m \frac{\beta}{\kappa^2r}AJ_m(\kappa r)-\frac{i}{\kappa}\omega \mu_0 BJ'_m(\kappa r), \\ 
        e_z^{(\nu)}(r) =& AJ_m(\kappa r),
    \end{split}
\end{equation*}
and for $r>r_f$
\begin{equation*}
    \begin{split}
        e_r^{(\nu)}(r) = \sum_{j=1,2}&\frac{i\beta}{q} C_j H'^{(j)}_m(q r) \\&- m\frac{\omega \mu_0}{q^2r}D_jH^{(j)}_m(q r), \\
        e_\phi^{(\nu)}(r) = \sum_{j=1,2}&-m\frac{\beta}{q^2r} C_j H^{(j)}_m(q r) \\&- \frac{i}{q}\omega \mu_0D_jH'^{(j)}_m(q r), \\
        e_z^{(\nu)}(r) = \sum_{j=1,2}&C_jH^{(j)}_m(q r).
    \end{split}
\end{equation*}
The functions $J_m(x)$ and $H_m^{(j)}(x)$ are the Bessel function of first kind and Hankel functions of $j$-th kind respectively, which takes the following form in the asymptotic limit ($|x| \gg 1$) 
\begin{equation*}
    \begin{split}
    J_m(x) &\approx \sqrt{\frac{2}{\pi x}}\cos{\left(x-\frac{m\pi}{2}-\frac{\pi}{4}\right)}, \\
    H^{(1)}_m(x) &\approx \sqrt{\frac{2}{\pi x}}e^{i\left(x-\frac{m\pi}{2}-\frac{\pi}{4}\right)}, \\
    H^{(2)}_m(x) &\approx \sqrt{\frac{2}{\pi x}}e^{-i\left(x-\frac{m\pi}{2}-\frac{\pi}{4}\right)}.
    \end{split}
\end{equation*} 
The prime denotes the derivative of the functions, $J'_m(x)=\frac{d}{dx}J_m(x)$ and $H'^{(j)}_m(x)=\frac{d}{dx}H_m^{(j)}(x)$. 

The coefficients $C_j$ and $D_j$ are related to the coefficients $A$ and $B$
\begin{equation*}
\begin{split}
    C_j &= (-1)^j\frac{i\pi q^2 r_f}{4}(AL_j+i\mu_0cBV_j), \\
    D_j &= (-1)^{j-1}\frac{i\pi q^2 r_f}{4}(i\epsilon_0cAV_j-BM_j),
\end{split}
\end{equation*}
where $\epsilon_0$ and $\mu_0$ is the vacuum permittivity and permeability respectively, and
\begin{equation*}
\begin{split}
    V_j &= \frac{mk\beta}{r_f\kappa^2q^2}(1-n_1^2)J_m(\kappa r_f)H_m^{(j)*}(qr_f), \\
    M_j &= \frac{1}{\kappa}J'_m(\kappa r_f)H_m^{(j)*}(qr_f)\\&-\frac{1}{q}J_m(\kappa r_f)H'^{(j)*}_m(qr_f), \\
    L_j &= \frac{n_1^2}{\kappa}J'_m(\kappa r_f)H_m^{(j)*}(qr_f) \\&-\frac{1}{q}J_m(\kappa r_f)H'^{(j)*}_m(qr_f).
\end{split}
\end{equation*}

$\mathbf{E}^{(\nu)}(\mathbf{r})$ are eigenvectors of the Hermitian operator $\mathcal{H} = [k^2-\frac{1}{\epsilon(\mathbf{r},\omega)}\nabla\times\nabla\times]$ with eigenvalues $\lambda_{(\nu)} = k_0^2-(\beta^2+q^2)$, where $k_0=\omega_0/c$ and $\omega_0$ is the argument frequency of the radiation Green's tensor. 

The profile function of a radiation mode is
\begin{equation*}
    \mathbf{e}^{(\nu)}(r) = e^{(\nu)}_r(r)\hat{\mathbf{r}}+e^{(\nu)}_\phi(r)\hat{\boldsymbol{\phi}}+e^{(\nu)}_z(r)\hat{\mathbf{z}},
\end{equation*}
and is normalized according to
\begin{equation*}
\begin{split}
    &\int_0^{2\pi}d\phi\int_0^{\infty}dr r\epsilon(\mathbf{r},\omega)\\ &\times[\mathbf{e}^{(\nu)}\cdot\mathbf{e}^{(\nu')*}]_{\beta=\beta',m=m'}=\delta_{ll'}\delta(\omega-\omega').
\end{split}
\end{equation*}

By choosing the constant $B=il\eta A$, the normalization of the profile function leads to the equations
\begin{equation*}
    \eta = \epsilon_0c(\frac{|V_j|^2+|L_j|^2}{|V_j|^2+|M_j|^2})^{1/2},
\end{equation*}
and 
\begin{equation*}
    1 = \frac{16\pi^2 k^2}{q^3}\left(|C_j|^2+\frac{\mu_0}{\epsilon_0}|D_j|^2\right).
\end{equation*}

\end{document}